\documentclass[conference]{IEEEtran}
\usepackage{cite}

\ifCLASSINFOpdf
	\usepackage[pdftex]{graphicx}
\else
  	\usepackage[dvips]{graphicx}
\fi
\graphicspath{{./_figures/},{./}}

\usepackage{amsmath}
\usepackage{xcolor}
\usepackage{siunitx}
\usepackage{textcomp}

\def\ie{i.e.,}

\begin{document}
%
\title{Winning versus losing during gambling \\and its neural correlates}


\author{
	\IEEEauthorblockN{%
		P.~Sacr\'{e}\IEEEauthorrefmark{1}, 
		M.~S.~D.~Kerr\IEEEauthorrefmark{1}, 
		S.~Subramanian\IEEEauthorrefmark{2}, 
		K.~Kahn\IEEEauthorrefmark{1},\\
		J.~Gonzalez-Martinez\IEEEauthorrefmark{3},
		M.~A.~Johnson\IEEEauthorrefmark{3}, 
		J.~T.~Gale\IEEEauthorrefmark{3}, and 
		S.~V.~Sarma\IEEEauthorrefmark{1}%
	}
	\IEEEauthorblockA{%
		\IEEEauthorrefmark{1}The Johns Hopkins University, Baltimore, MD.
		\IEEEauthorrefmark{2}University of Cambridge, UK.
		\IEEEauthorrefmark{3}Cleveland Clinic, Cleveland, OH.
	}
}


%


\maketitle

\begin{abstract}
	Humans often make decisions which maximize an internal utility function. For example, humans often maximize their expected reward when gambling and this is considered as a ``rational'' decision. However, humans tend to change their betting strategies depending on how they ``feel''. If someone has experienced a losing streak, they may ``feel'' that they are more likely to win on the next hand even though the odds of the game have not changed. That is, their decisions are driven by their emotional state. 
	In this paper, we investigate how the human brain responds to wins and losses during gambling. 
	Using a combination of local field potential recordings in human subjects performing a financial decision-making task, spectral analyses, and non-parametric cluster statistics, we investigated whether neural responses in different cognitive and limbic brain areas differ between wins and losses after decisions are made.
	In eleven subjects, the neural activity modulated significantly between win and loss trials in one brain region: the anterior insula~($p=0.01$). In particular, gamma activity (\SIrange{30}{70}{\hertz}) increased  in the anterior insula when subjects just realized that they won. Modulation of metabolic activity in the anterior insula has been observed previously in functional magnetic resonance imaging studies during decision making and when emotions are elicited. However, our study is able to characterize temporal dynamics of electrical activity in this brain region at the millisecond resolution while decisions are made and after outcomes are revealed. 
\end{abstract}


%
\IEEEpeerreviewmaketitle

\section{Introduction}
Decision-making links cognition to behavior and is a key driver of human personality, fundamental for survival, and essential for our ability to learn and adapt. It has been well-established that humans often make emotional-based decisions~\cite{Toda:1980aa}. Thus, psychiatric patients who have dysfunctional cognitive circuitry, frequently have alterations in decision-making that are poorly understood. 
	
Understanding the neural basis of decision-making is therefore essential toward patient management. However, access to the human brain has been limited to a few case studies wherein subjects have lesions in the key decision-making structures such as the orbital frontal cortex~\cite{Bechara:2004ab,Kim:2005aa,Bohn:2003aa}, or wherein functional magnetic resonance imaging (fMRI) is used to measure neural activity in several healthy subjects during decision-making~\cite{Logothetis:2008aa}. Both of these approaches have limitations. Lesions don't provide actual neural data to ascertain a specific brain regions's role during behavior. Rather, the region's function is inferred by absence of behaviors from lesioned subjects when compared to healthy subjects. On the other hand, fMRI provides a correlate of neural activity (metabolic activity) but suffers from poor temporal resolution. fMRI resolution is on the order of multiple seconds, while decisions are often made on the order of tenths of a second.
		
Here, we took advantage of a technique called stereoelectroencephalography~(SEEG) that allowed us to record high temporal resolution electrophysiological data (electrical activity at the millisecond scale) directly from deep and peripheral brain regions in human subjects while they performed a gambling task. These subjects are implanted with electrodes for clinical purposes, and each contact in the brain generates a local field potential (LFP) signal.
	
For our gambling task, eleven subjects played a game of high card where they won virtual money if their card was higher than the computer's card.  On each trial, subjects had to decide to bet ``high'' (\SI{20}[\$]{}) or ``low'' (\SI{5}[\$]{}) on their card being higher than the hidden computer's card. Eighty percent of the trials lead to clear rational decisions where the expected reward for one choice is higher than another. On twenty percent of the trials, however, there is no clear rational decision since the expected reward is equal for both choices. Here, we focused on characterizing differences in brain activity between when a subject realizes that he/she has won versus when he/she has lost. Thus, the goal of this study is to assess the role of different brain regions in responding to outcomes of gambling decisions.
	
To identify neural correlates that encode a subject's response to a win versus a loss, we computed spectrograms for each brain region (\ie{} electrode channel) and each patient across all trials when subjects won and compared them to spectrograms when subjects lost. Specifically, we examined spectral content before, during, and after the computer's card was shown (and hence outcome is revealed) and implemented a non-parametric cluster statistic to test whether the spectral activity differed between wins and losses in the given brain region. The non-parametric cluster test generates a cluster defined by a set of adjacent time-frequency windows that gives rise to the smallest \mbox{$p$-value}. If this \mbox{$p$-value} is smaller than 0.05, then the brain region (defined by location of the channel) was designated as encoding response to gambling outcome.
	
We found that upon examining several brain regions in cognitive, limbic and hippocampal networks, the anterior insula encoded information that separated wins from losses. Specifically, when gamma activity (\SIrange{30}{70}{\hertz}), was prevalent in this region, then the player was more likely to have won after he/she sees the computer's card. The anterior insula has been implicated in the role of emotions in risk-related decision making~\cite{Sanfey:2003aa} as well as being a neural basis for encoding how someone ``feels''~\cite{Craig:2009aa}. But no study has had access to electrical activity from these regions at a millisecond time-scale. Hence, these findings are the first to show evidence of temporal dynamics of rhythms in the brain at a fine resolution when decisions are made and when subjects ``feel'' the consequences of their decisions. 





\section{Methods} 
\label{sec:methods}

\subsection{Subjects} 
\label{sub:subjects}


At the Cleveland Clinic, patients with medically intractable epilepsy routinely undergo SEEG recordings in order to localize the seizure focus. See~\ref{sub:stereoelectroencephalography} for details on the SEEG procedure. In this study, aside from the behavioral experiments, no alterations were made to the patient's clinical care, including the placement of the electrodes~\cite{Johnson:2014aa}.  Subjects enrolled voluntarily and gave informed consent under criterion approved by the Cleveland Clinic Institutional Review Board. A total of eleven subjects volunteered to perform the task. Details on these recordings and eventual annotated seizure focus of these eleven patients are noted in Table~\ref{tab:subjects}. 

Subjects were implanted with 8 to 13 depth electrodes.  Implantation was performed using robot-assisted surgery along with co-registered functional MRIs and angiograms to ensure safe implantation~\cite{Ongur:2000aa}.  Once inserted, SEEG electrophysiological data were acquired using a Nihon Kohden 1200 EEG diagnostic and monitoring system (Nihon Kohden America, USA) at a sampling rate of \SI{2}{\kilo\hertz}.  Behavioral event data were simultaneously acquired through the MonkeyLogic MATLAB$^\text{\textregistered}$ toolbox~\cite{Asaad:2008aa}.

There are standard concerns in analyzing data from epileptic patients.  First, patients are often on medication, which might affect the neurophysiology of the brain.  For clinical purposes, patients were kept off of their anti-seizure medication for their entire stay at Cleveland Clinic, so these effects would be minimized.  Secondly, actual seizures might impact the neurophysiology around the seizure focus. Human epilepsy recordings are taken to localize the seizure focus, so overlap is expected between seizure focus and areas recorded.

\renewcommand{\extrarowheight}{5pt}
\begin{table}[tbp]
	\centering
	\caption{%
		This table provides clinically relevant information on each subject: the gender, the age in years, and the duration of epilepsy (``Dur.'') in years.
		This table also provides the number of win, loss and draw trials.  
	}
	\begin{tabular}{clrrrrr}
		\textit{ID} & \textit{Gender} & 
		\begin{tabular}[t]{@{}r@{}}\textit{Age }\\[-5pt] {\scriptsize\emph{[yr.]}}\end{tabular} & 
		\begin{tabular}[t]{@{}r@{}}\textit{Dur.}\\[-5pt] {\scriptsize\emph{[yr.]}}\end{tabular} & 
		\textit{Win} & \textit{Loss} & \textit{Draw} \\
		1           & male            & 26                   & 3                         & 73           & 73            & 39            \\
		2           & female          & 41                   & 38                        & 79           & 58            & 25            \\
		3           & female          & 55                   & 52                        & 23           & 34            & 15            \\
		4           & female          & 31                   & 13                        & 53           & 57            & 34            \\
		5           & female          & 53                   & 23                        & 50           & 61            & 25            \\
		6           & female          & 60                   & 8                         & 63           & 88            & 21            \\
		7           & female          & 36                   & 36                        & 56           & 68            & 33            \\
		8           & female          & 23                   & 5                         & 46           & 50            & 36            \\
		9           & male            & 32                   & 11                        & 65           & 59            & 36            \\
		10          & female          & 32                   & 13                        & 57           & 62            & 35            \\
		11          & male            & 28                   & 11                        & 67           & 78            & 37           
	\end{tabular}
	\label{tab:subjects}
\end{table}%


\subsection{Stereoelectroencephalography} 
\label{sub:stereoelectroencephalography}

The Cleveland Clinic is a world-renowned center for the evaluation and treatment of epilepsy and brain tumors, assessing around \si{9500} patients every year from all \si{50} states and more than \si{10} countries. More than \si{400} epilepsy surgeries are performed every year, including a growing number of stereotactically implanted depth electrodes or stereoelectroencephalography (SEEG)~\cite{Gonzalez-Martinez:2014ac,Gonzalez-Martinez:2014ab,Gonzalez-Martinez:2014ad,Gonzalez-Martinez:2014aa}. This surgical procedure was developed in France~\cite{Bancaud:1976aa,Bancaud:1975aa,Bancaud:1966aa,Bancaud:1974aa,Bancaud:1968aa,Bancaud:1969aa,Talairach:1965aa,Talairach:1980aa,Talairach:1992aa,Bancaud:1965aa,Munari:1979aa,Munari:1982aa,Munari:1989aa}, and brought to the United States by Dr.~Jorge~Gonzalez-Martinez~\cite{Gonzalez-Martinez:2013aa}. The innovative approach using SEEG methodology relies on its capability in accessing large-scale networks, providing precise human brain data, from cortical to subcortical areas, in three-dimensional fashion. In routine placement of depth electrodes, burr-holes that are each \SI{15}{\milli\meter} in diameter are required for safe visualization of cortical vessels, and therefore only a small number of electrodes are placed. SEEG placement, however, uses several small drill holes (\SI{1.8}{\milli\meter} in diameter), allowing many electrodes to be inserted. 

\begin{figure}[tbp]
	\centering
	\includegraphics[width=\columnwidth]{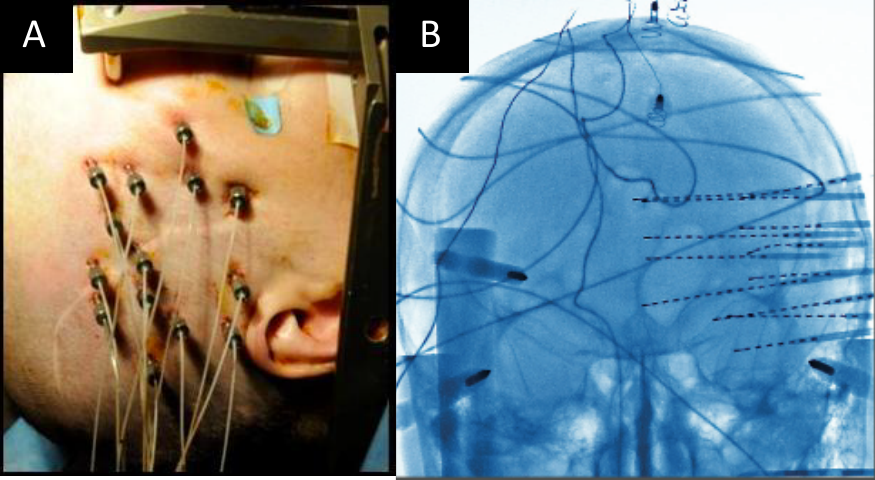}
	\caption{%
		Imaging fusion and placement of multiple electrodes using the SEEG method. 
		Fig. A is a photograph showing 14 electrodes at the skin surface. 
		Fig. B is a fluoroscopy image of an SEEG-implanted subject (coronal view with eye forward). Note the precise parallel placement, with tips terminating at the midline or dural surface.
		}
	\label{fig:seeg}
\end{figure}

Since direct visualization of the cortical surface is not possible with small drills (Fig.~\ref{fig:seeg}A), the SEEG technique requires detailed pre-procedural vascular mapping using pre-operative imaging with magnetic resonance angiography (MRA) and cerebral angiography. Angiography is an X-ray examination of the blood vessels. The mapping procedure is performed under fluoroscopy using general anesthesia, and an expert neuro-anesthesiologist correctly titrates anesthesia to permit measurement of intracranial electroencephalography (EEG). The number and location of implanted electrodes are pre-operatively planned based on a pre-implantation hypothesis, which is formulated in accordance with non-invasive pre-implantation data, such as seizure semiology, ictal and inter-ictal scalp EEG, magnetic resonance imaging (MRI), positron emission tomography (PET) and ictal single-photon emission computed tomography (SPECT) scans. Thus, the implantation strategy has the goal of accepting or rejecting the pre-implantation hypothesis of the location of the epileptogenic zone (EZ). 

SEEG provides a more complete coverage of the brain, from lateral, intermediate and/or deep structures in a three-dimensional arrangement recorded over hundreds of channels. Using strict techniques, this procedure is safe and minimally invasive: only 1/1176 implantations last year resulted in an asymptomatic intracranial hemorrhage. The rate of complications in SEEG implantations is less than \SI{1}{\%}~\cite{Gonzalez-Martinez:2014ad,Cardinale:2013aa}.


\subsection{Gambling task} 
\label{sub:gambling_task}

Subjects performed the gambling task in their Epilepsy Monitoring Unit room.  The task was displayed via a computer screen and the subject interacted with the task using an InMotion2 robotic manipulandum (Interactive Motion Technologies, USA). The manipulandum is controlled by the subject's hand and allows for 2D planar motion, which translated directly to the position of a cursor on screen.  

\begin{figure}[tbp]
	\centering
	\includegraphics{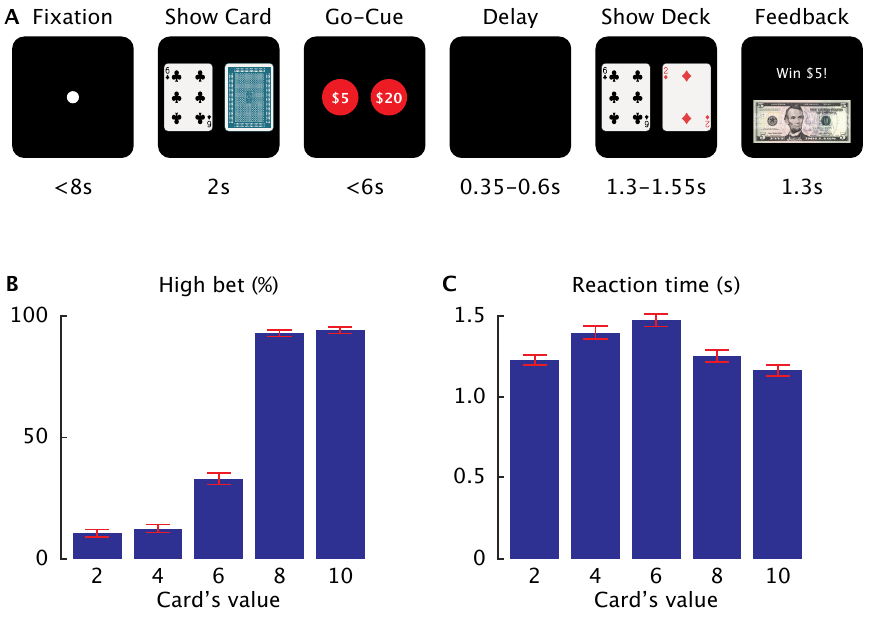}
	\caption{%
		Gambling task and behavioral results. 
		(A)~Timeline of the behavioral task.  After fixation, subjects were shown their card.  Once the bets were shown, subjects selected one of the choices and then were shown the computer's card following a delay. Feedback was provided afterwards by displaying the amount won or lost.
		(B)~Average bet decisions across cards.  Subjects predominantly bet low for 2 and 4 cards and bet high for 8 and 10 cards.  There was no predominant strategy for 6 cards, which had bout \SI{33}{\%} chance of eliciting a high bet.
		(C)~Reaction times across cards. Subjects reacted faster for cards whose rewards had lower variability. }
	\label{task-fig}
\end{figure}

The gambling task (Fig.~\ref{task-fig}A) is based on a simple game of high card where subjects would win virtual money if their card beat the computer's card.
Specifically, in the beginning of each trial, the subject controlled a cursor via a planar manipulandum to a fixation target. 
Afterwards, the subject is shown his card (2, 4, 6, 8, or 10) that is randomly chosen with equal distribution.  The computer's card is initially hidden. 
The screen then shows their two choices: a high bet (\SI{20}[\$]{}) or a low bet (\SI{5}[\$]{}). The subject has \si{6} seconds to select one with his cursor. 
Following selection, the computer's card, which follows the same distribution, is revealed.
The final screen depicts the amount won or lost.  

\subsection{Data analysis} 
\label{sub:data_analysis}

All electrophysiological and behavioral analyses were conducted offline using custom MATLAB$^\text{\textregistered}$ scripts. 

Data for 16 anatomical regions in cognitive, limbic and hippocampal networks were separated into win and loss trials. For each brain region, differences in the neural responses between the task conditions during the \SI{250}{\milli\second} before and \SI{1000}{\milli\second} after the computer card were examined by means of a non-parametric cluster statistic.
Specifically, spectrograms were constructed for each trial time-locked to when the computer's card is shown. 
Then the spectrograms for win trials were compared to those for loss trials. To see if spectrograms for each group were statistically significantly different, we used a nonparametric cluster-based test~\cite{Maris:2007aa}. Clusters are defined as a set of adjacent time-frequency windows whose activity is statistically significant between trials where the subject ends up winning versus losing.  


\subsubsection{Spectral analysis} 
\label{ssub:spectral_analysis}

We calculated the oscillatory power using multitapers from the Chronux toolbox~\cite{Bokil:2010aa}. We used three orthogonal tapers with a \SI{300}{\milli\second} window sliding at \SI{50}{\milli\second} steps. We dropped frequencies under \SI{10}{\hertz} because of the Nyquist criterion and analyzed upwards to \SI{100}{\hertz}. Afterwards, we normalize each frequency bin's power by first taking the natural log of the power in each frequency bin, and then performing a Gaussian normalization based on the power in each frequency bin over the entire recording session. 


\subsubsection{Non-parametric cluster statistical test} 
\label{ssub:nonparametric_statistical_test}
Significant differences between the neural response data in each anatomical region are defined by a non-parametric cluster statistic run on data aggregated from trials by all relevant subjects~\cite{Maris:2007aa}.

This test leverages the dependency between adjacent time-frequency windows in order to avoid over-penalizing with multiple comparison corrections. For each time-frequency window in the spectrogram, a null distribution was created by shuffling these wins and loss labels 1000 times between trials within each subject. Within each shuffle, the average difference between the newly labeled win and loss trial spectrograms was calculated. A \mbox{$p$-value} was assigned for each window by comparing the difference acquired from the true labels with the distribution of differences acquired from the shuffled labels.  Clusters were formed by grouping windows with significant \mbox{$p$-values} ($p<0.05$) that were adjacent in either time or frequency.  The test statistic for each cluster was calculated by taking the sum of the log of the \mbox{$p$-values} for each window in the cluster.  This prioritizes clusters that both have strong differences as well as large sizes.  A null distribution of cluster statistics was created using the same process but with the 1000 spectrograms obtained from the originally shuffled labels.  The observed cluster statistic was then compared against this null distribution of cluster statistics in order to obtain the final \mbox{$p$-value} of the test. 


%




\section{Results} 
\label{sec:results}

This section summarizes the main findings of our analysis. 
First, we show the \mbox{$p$-value} associated with 16 brain regions. 
Then, we focus on the brain region identified in our exploratory analysis and show time-frequency differences of the neural responses around the show computer card epoch.

\begin{figure}[tbp]
	\centering
	\includegraphics{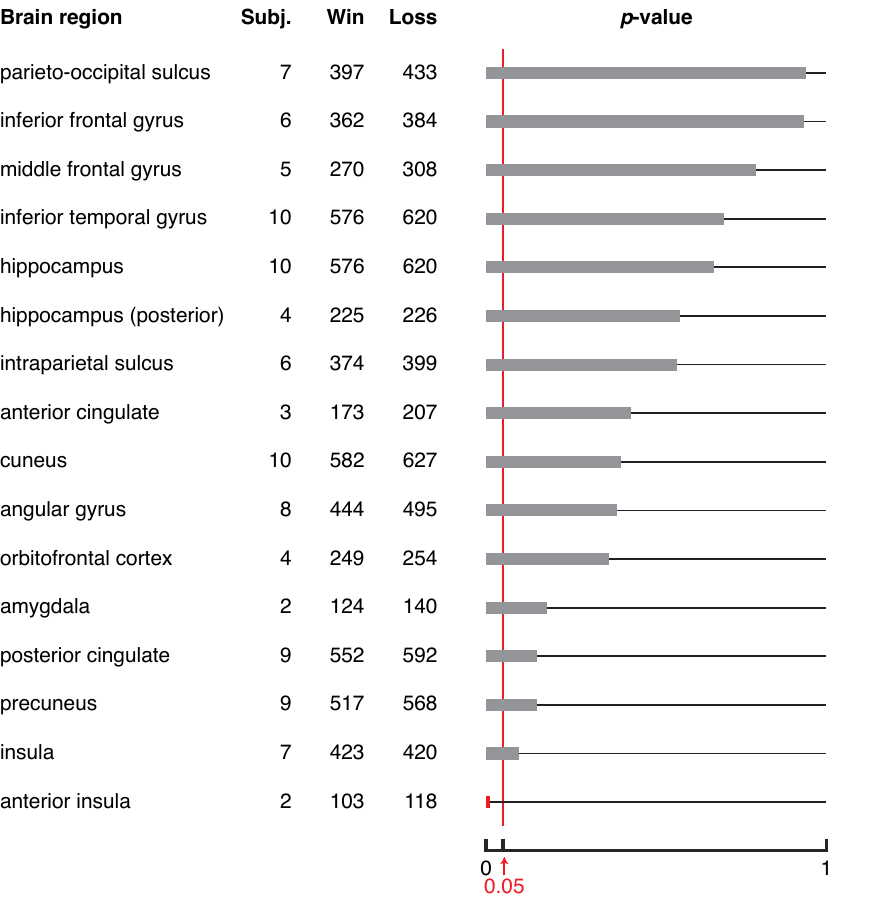}
	\caption{%
		Non-parametric cluster statistic for different brain regions. 
		For each brain region, differences in the neural responses between the task conditions during the \SI{250}{\milli\second} before and \SI{1000}{\milli\second} after the show computer card epoch were summarized by means of a non-parametric cluster statistic (\mbox{$p$-value}). 
		In addition, we provide the number of subjects with recording in this area~(Subj.), and the number of win and loss trials.
	}
	\label{fig:irr-rat-behavior}
\end{figure}

While multiple brain regions appear to show some response (data not shown), the two task conditions only significantly differ after the computer card is shown in one brain region: anterior insula~($p=0.01$) (see Fig.~\ref{fig:irr-rat-behavior}). All \mbox{$p$-values} are computed from a non-parametric cluster statistic described in Section~\ref{sec:methods}. 

\begin{figure}[tbp]
  	\centering
    \includegraphics{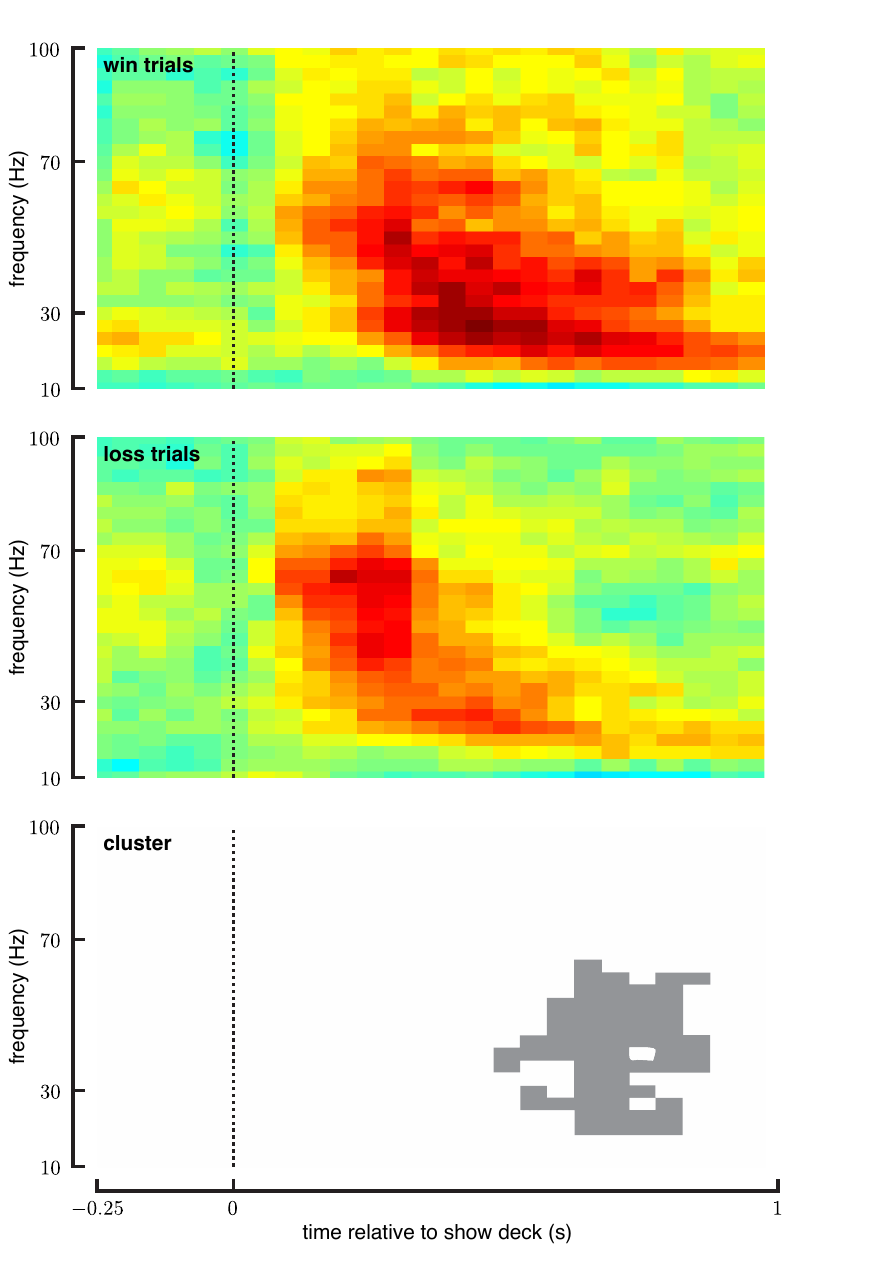}
  	\caption{%
		Differences in the neural responses during the \SI{250}{\milli\second} before and \SI{1000}{\milli\second} after the show computer card epoch for anterior insula. 
		(First and second rows)~Spectrograms of the neural responses show differences in the time-frequency domain between win and loss conditions.
		(Third row)~The cluster emphasizes the region of the time-frequency domain where the neural responses show significant differences between win and loss conditions ($p < 0.05$).
	}
  	\label{fig:cluster}
\end{figure}

Differences in the neural responses around the show computer card epoch was examined for the anterior insula (Fig.~\ref{fig:cluster}). Spectrograms of the neural responses between win and loss conditions show large differences in the gamma band (\SIrange{30}{70}{\hertz}) after the computer card is shown and the gambling outcome is revealed.

\begin{figure}[tbp]
  	\centering
    \includegraphics{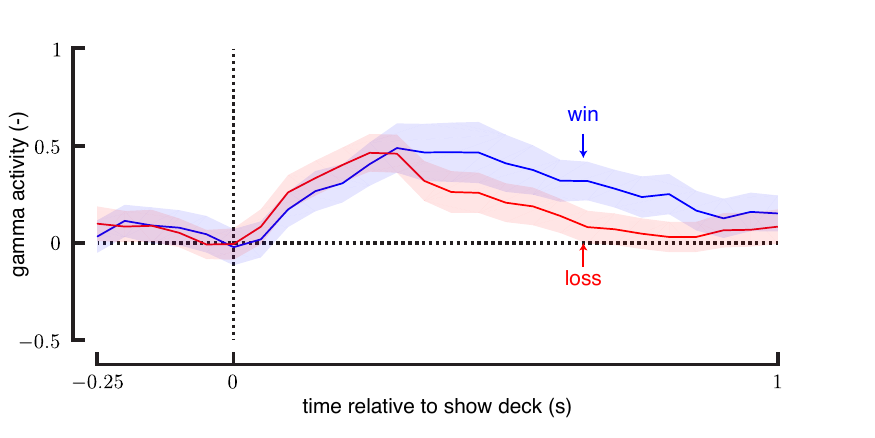}
  	\caption{
		Gamma activity (\SIrange{30}{70}{\hertz}) increases in anterior insula quickly around \SI{250}{\milli\second} after the computer card epoch and then decreases slowly until \SI{1000}{\milli\second} after the computer card epoch. The gamma activity for win trials decreases more slowly than gamma activity for loss trials.		
		}
	\label{fig:power}
\end{figure}

We summarize the neural activity in the time-frequency domain by averaging the responses over \SIrange{70}{100}{\hertz}. Gamma activity increases in anterior insula quickly around \SI{250}{\milli\second} after the computer card epoch and then decreases slowly until \SI{1000}{\milli\second} after the computer card epoch (Fig.~\ref{fig:power}). The gamma activity for win trials decreases more slowly than gamma activity for loss trials.


\section{Conclusion} 
\label{sec:conclusion}
In our subject population, the neural activity modulated significantly between win and loss trials in the anterior insula~($p=0.01$). In particular, gamma-band activity increased around \SI{500}{ms} after the show computer card for win trials. 
The anterior insula has been implicated in the role of emotions in risk-related decision making~\cite{Sanfey:2003aa} as well as being a neural basis for encoding how someone ``feels''~\cite{Craig:2009aa}, but no study has had access to electrical activity from these regions at a millisecond time-scale. Hence, these findings are the first to show evidence of temporal dynamics of rhythms in the brain at a fine resolution when decisions are made and when subjects ``feel'' the consequences of their decisions. Still, the exact mechanisms by which this region in involved in risk-based decision making are not known and will therefore remain the focus of future work.





\section*{Acknowledgment}

The authors would like to thank J.~Bulacio, J.~Jones, \mbox{H.-J.~Park} and S.~Thompson (Cleveland Clinic, Cleveland, OH) for facilitating experiments, collecting, de-identifying and transferring data, and providing anatomical labels.
This work was supported by a National Science Foundation grant  (EFRI-MC3: \# 1137237) awarded to S.V.S., J.A.G., J.B. and J.T.G.



%
%
%

\bibliographystyle{ieeetr}
\bibliography{%
	/Users/pierresacre/Work/Research/biblio/biblio_sacre/journal-name-abbrv,%
	/Users/pierresacre/Work/Research/biblio/biblio_sacre/bibfile%
}

\end{document}